\documentclass[%
 reprint,
 superscriptaddress,
 showpacs,
 preprintnumbers,
 amsmath,amssymb,
 aps,
 prd,
 floatfix,
]{revtex4-1}

\usepackage{dcolumn}
\usepackage{bm}
\usepackage{hyperref}

\usepackage{graphicx, psfrag, epstopdf}
\graphicspath{{fig/}}

\def\<{\left\langle}
\def\>{\right\rangle}

\def\opsi{\overline{\psi}}

\def\hmu{\hat{\mu}}

\def\mua{\mu a}
\def\dmu{\partial (\mu a)}

\newcommand{\beq} {\begin{eqnarray}}
\newcommand{\eeq} {\end{eqnarray}}

\newcommand{\tr}{ {\rm Tr} \: }

\begin{document}

\title{Study of the phase diagram of dense two-color QCD within lattice simulation}

\author{V.~V.~Braguta}
\email[]{braguta@itep.ru}
\affiliation{Institute for High Energy Physics NRC "Kurchatov Institute", Protvino, 142281 Russia}
\affiliation{Institute for Theoretical and Experimental Physics NRC "Kurchatov Institute", Moscow, 117218 Russia}
\affiliation{School of Biomedicine, Far Eastern Federal University, Sukhanova 8, Vladivostok, 690950 Russia}
\affiliation{Moscow Institute of Physics and Technology, Institutskii per. 9, Dolgoprudny, Moscow Region, 141700 Russia}

\author{E.-M. Ilgenfritz}
\email[]{ilgenfri@theor.jinr.ru}
\affiliation{Joint Institute for Nuclear Research, BLTP, Dubna, 141980 Russia}

\author{A.~Yu.~Kotov}
\email[]{kotov@itep.ru}
\affiliation{Institute for Theoretical and Experimental Physics NRC "Kurchatov Institute", Moscow, 117218 Russia}
\affiliation{National Research Nuclear University MEPhI (Moscow Engineering Physics Institute), Kashirskoe Highway, 31, Moscow 115409, Russia}

\author{A. V. Molochkov}
\email[]{molochkov.av@dvfu.ru}
\affiliation{School of Biomedicine, Far Eastern Federal University, Sukhanova 8, Vladivostok, 690950 Russia}

\author{A.~A.~Nikolaev}
\email[]{nikolaev.aa@dvfu.ru}
\affiliation{School of Biomedicine, Far Eastern Federal University, Sukhanova 8, Vladivostok, 690950 Russia}
\affiliation{Institute for Theoretical and Experimental Physics NRC "Kurchatov Institute", Moscow, 117218 Russia}

\date{\today}

\begin{abstract}
In this paper we carry out a low-temperature scan of the phase diagram of dense two-color QCD with 
$N_f=2$ quarks. 
The study is conducted using
lattice simulation with rooted staggered quarks. At small chemical
potential we observe the hadronic phase, where the theory is in a confining state,
chiral symmetry is broken, the baryon density is zero and there is no diquark condensate.
At the critical point $\mu = m_{\pi}/2$ we observe the expected second order transition to 
Bose-Einstein condensation of scalar diquarks. 
In this phase the system is still in confinement in conjunction with non-zero baryon density, but the chiral symmetry is restored
in the chiral limit. We have also found that in the first two phases the system is well described by chiral perturbation 
theory. For larger values of the chemical potential the system turns into another phase, where 
the relevant degrees of freedom are fermions residing inside the Fermi sphere, 
and the diquark condensation takes place
on the Fermi surface. In this phase the system is still in confinement, chiral symmetry is restored and 
the system is very similar to the quarkyonic state predicted by SU($N_c$) theory at large $N_c$.
\end{abstract}



\maketitle

\section{Introduction.}

The phase  diagram of QCD is of high importance for several fields of
observational physics like cosmology and astrophysics. One field of
experimental physics, located between nuclear physics and high energy
physics, is the study of hadronic matter created by relativistic heavy 
ion collisions. Such experiments are 
addressing the structure of the phase 
diagram, although the understanding and modeling of an actual collision
requires much more than the knowledge of the equilibrium phase diagram.
However, equilibrium observables like the equation of state and 
transport coefficients are highly needed to be used in hydrodynamical 
approaches which serve to probe various scenarios.

The region of high temperature and vanishing baryonic density 
of QCD phase diagram is well explored at LHC and RHIC. The theoretical 
study of this part of the phase diagram can be carried out with lattice 
gauge theory (LGT), based on the first principles of QCD. Today this approach 
has reached a high level of maturity and the results obtained within LGT for
small $\mu/T$ are in good agreement with experiments~\cite{Ding:2015ona, Philipsen:2005mj}. 

In the 2010-th years, a ``low-energy frontier'' of heavy ion physics has 
opened (with the beam energy scan program at RHIC) focussing at the region 
of high baryonic density and lower temperatures. The new experimental
facilities presently under construction, FAIR and NICA, hosting the future 
experiments CBM, BM\&N, and MPD, respectively, will be suitable for this 
region of the phase diagram. This situation is urging theorists to study 
QCD with large chemical potential. 

Unfortunately, lattice simulation of QCD cannot be applied today to arbitrary 
chemical potential because of the sign problem~\cite{Muroya:2003qs}. The origin
of the sign problem is that the fermion determinant becomes complex-valued, 
and direct simulation by importance sampling of gauge field configurations
is not possible. 
In the absence of straightforward results from LGT one applies different 
approaches to study the $(T,\mu)$ phase diagram: for instance, mean field 
approaches~\cite{Buballa:2003qv}, the method of Dyson-Schwinger equations~\cite{Hong:1999fh}, 
the large--$N_c$ approach~\cite{McLerran:2007qj, Hanada:2011ju}, perturbative QCD coupled to
HRG models~\cite{Gorda:2014vga}, exploring the phase diagram of QCD with isospin chemical potential~\cite{Son:2000xc, Kogut:2002zg, Kogut:2004zg, deForcrand:2007uz, Endrodi:2014lja}
and others. Although the results obtained within these approaches are very interesting,
they may still be rather schematic and require confirmation. 

An alternative to lattice simulation of SU(3) QCD with $\mu \neq 0$ is the 
simulation of SU(2) QCD (also called QC$_2$D). Introduction of a chemical 
potential to the latter theory does not lead to a sign problem, so one can 
apply the standard lattice approach to study this theory. Although a two-color 
world differs from the tree-color world, lattice study of QC$_2$D with chemical 
potential can provide us with important information about the properties of QCD 
with non-zero baryon density. In particular we believe that some physical properties of the regions 
of the phase diargam where relevent degrees of freedom are quarks and gluons
are similar for the SU(2) and SU(3) theories~\cite{Hanada:2011ju}. As an example one could 
mention equation of state, some properties of gluon propagator (for instance, Debye screening), 
generation of the fermion mass gap, etc. In addition one can use 
SU(2) QCD to study how non-zero density influences different observables and 
phenomena. We would like also to note that the QC$_2$D phase diagram has a rich
structure and it is interesting to study by its own.

The properties of QC$_2$D were studied theoretically within the following 
approaches: chiral perturbation theory
(ChPT)~\cite{Kogut:1999iv, Splittorff:2000mm, Kogut:2000ek, Splittorff:2001fy, Kanazawa:2009ks},
the NJL model~\cite{Ratti:2004ra, Brauner:2009gu, Sun:2007fc, He:2010nb},
functional renormalization group~\cite{Strodthoff:2011tz, Strodthoff:2013cua, vonSmekal:2012vx, Kamikado:2012bt},
random matrix theory~\cite{Vanderheyden:2001gx, Kanazawa:2009en, Kanazawa:2011tt}. 
Principally, these studies have revealed the following phase structure 
of low temperature QC$_2$D with three subsequent phases:
(1) $0 < \mu < \mu^c$ (hadronic phase), 
(2) $\mu^c < \mu < \mu^d$ (``baryon onset'' with a superfluid condensate 
due to Bose-Einstein condensation [BEC]) 
and (3) $\mu^d < \mu$ (the phase with diquark condensation due to the 
Bardeen-Cooper-Schrieffer mechanism [BCS]~\cite{Kanazawa:2012zr}).

The first lattice study of 
QC$_2$D with chemical potential and Wilson fermions was performed by 
A. Nakamura in~\cite{Nakamura:1984uz}. 
Futher lattice investigation of dense two-color QCD was continued by
J. Kogut and collaborators~\cite{Hands:1999md} using staggered quarks. The staggered Dirac operator
without rooting describes $N_f=4$ flavors. Making the whole fermion matrix hermitian positive definite
by doubling the number of flavors has lead to the eight-flavor theory investigated in the pioneering
paper~\cite{Hands:1999md}. Following this work, introduction of the rooting trick for the staggered
fermion determinant has allowed to investigate the case of $N_f=4$ flavors in more detail~\cite{Kogut:2001if,
Kogut:2001na, Kogut:2002cm}. 
The main activity in two-color QCD was later continued by the Swansea group (S. Hands and collaborators),
mainly for the two-flavor theory, with each flavor described by one species of Wilson fermions~\cite{Hands:2006ve, Hands:2010gd, Cotter:2012mb, Boz:2013rca}.

In this paper we are going to study the QC$_2$D phase diagram with $N_f=2$ 
flavors going back to the lattice simulation of staggered fermions using the 
rooting procedure. The advantage of the staggered fermion formulation is the
approximate residual chiral symmetry of the Dirac operator~\cite{Montvay}. Therefore we have
chosen this formulation to complement the Swansea studies by an alternative study of the 
two flavors case. In our first paper~\cite{Braguta:2015cta} we have calculated the Polyakov 
loop and the chiral condensate as functions of 
temperature for different values
of the chemical potential $\mu$\footnote{Quark chemical potential is understood by $\mu$ here
and below.}. In the present paper we are going to carry out a $\mu$ scan at low temperature
of the QC$_2$D phase diagram.

The paper is organized as follows. In sect. II we specify details of the lattice set-up to be used: action, 
the way of the diquark source introduction, details of the simulation. In sect. III we present the numerical
results of this study. The last section is devoted to the discussion of the results and to some conclusions
to be drawn.

\section{The lattice set-up}

\subsection{Partition function}

In our simulations we used the Wilson action for the SU(2) gauge fields
\beq
\label{eq:S_G}
S_G = \beta \sum_{x}\sum_{\mu < \nu = 1}^4 \Bigl(1 - \frac{1}{2} \tr U_{x, \mu\nu} \Bigr).
\eeq
For the fermionic degrees of freedom we used staggered fermions with an action of the form
\begin{equation}
\label{eq:S_F}
S_F = \sum_{x, y} \opsi_x M(\mu, m)_{x, y} \psi_y + \frac{\lambda}{2} \sum_{x} \left( \psi_x^T \tau_2 \psi_x + \opsi_x \tau_2 \opsi_x^T \right)\,,
\end{equation}
\beq
\label{eq:Dirac_operator}
M_{xy} = ma\delta_{xy} &+& \frac{1}{2}\sum_{\mu = 1}^4 \eta_\mu(x)\Bigl[ U_{x, \mu}\delta_{x + \hmu, y}e^{\mua\delta_{\mu, 4}} \nonumber \\
&-& U^\dagger_{x - \hmu, \mu}\delta_{x - \hmu, y}e^{- \mua\delta_{\mu, 4}} \Bigr].
\eeq
where $\opsi$, $\psi$ are staggered fermion fields, $a$ is the lattice spacing, $m$ is the bare quark mass,
and $\eta_\mu(x)$ are the standard staggered phase factors: $\eta_1(x) = 1,\, \eta_\mu(x) = (-1)^{x_1 + ...+ x_{\mu-1}},~\mu=2,3,4$.
The chemical potential $\mu$ is introduced into equation ~\eqref{eq:Dirac_operator} through the multiplication of the links along
and opposite 
to the temporal direction by factors $e^{\pm \mua}$ respectively. 
This way of introducing
the chemical potential makes it possible to avoid additional divergences and to reproduce well known
continuum results~\cite{Hasenfratz:1983ba}.

In addition to the standard staggered fermion action we add 
a diquark source term~\cite{Hands:1999md}
to equation (\ref{eq:S_F}). The diquark source term explicitly violates $U_V(1)$ and allows to observe diquark 
condensation even on finite lattices,
because this term effectively chooses one vacuum from the family of $U_V(1)$-symmetric vacuums. 
The results presented in this paper are obtained as follows: we carry out 
simulations at small but nonzero parameter $\lambda \ll ma$, and then extrapolate obtained data to $\lambda \to 0$. Notice that similar to the diquark source term 
an additional pion term was introduced to the fermion action during the studies of QCD phase diagram with isospin chemical potential~\cite{Kogut:2002zg, Kogut:2004zg, Endrodi:2014lja}.

Integrating out the fermion fields the partition function for the theory with the action $S=S_G+S_F$
can be written in the form
\beq
Z &=& \int DU e^{-S_G} \cdot Pf \begin{pmatrix} \lambda \tau_2 & M \\ -M^T & \lambda \tau_2 \end{pmatrix} \nonumber \\
&=& \int DU e^{-S_G} \cdot {\bigl ( \det (M^\dagger M + \lambda^2) \bigr )}^{\frac 1 2},
\label{z1}
\eeq
which corresponds to $N_f=4$ dynamical fermions in the continuum limit. Note that the pfaffian $Pf$
is strictly positive, such that 
one can use 
Hybrid Monte-Carlo methods to study this system.
The lattice study of the theory with partition function (\ref{z1}) was carried out in papers
\cite{Kogut:2001na, Kogut:2001if, Kogut:2002cm}. In 
the present paper we are going to study the theory with the partition function
\beq
Z=\int DU e^{-S_G} \cdot {\bigl ( \det (M^\dagger M + \lambda^2) \bigr )}^{\frac 1 4},
\label{z2}
\eeq
which corresponds to $N_f=2$ dynamical fermions in the continuum limit.
Notice that the diquark source term lifts the lowest eigenvalues 
of the matrix in determinant (\ref{z2}) and thus lowers the cost of numerical simulations. 

It is known that the symmetries of the staggered fermion action are different from those of two-color QCD
with fundamental quarks \cite{Hands:1999md}. In particular, the symmetry breaking pattern of QC$_2$D
with fundamental quarks is SU(2$N_f$) $\to$ Sp(2$N_f$), whereas for staggered quarks it is SU(2$N_f$) $\to$ O(2$N_f$). However, it is easy to show that the diquark source term in the continuum
limit can be written as
\beq
\frac{\lambda}{2} \sum_{x} \left( \psi_x^T \tau_2 \psi_x + \opsi_x \tau_2 \opsi_x^T \right) \biggr|_{a \to 0} = \nonumber \\ \frac{\lambda}{2} \int d^4 x~  \bigl(  q_i^T  C \gamma_5 \tau_2  q_j + \bar q_i  C \gamma_5 \tau_2  \bar q_j^T \bigr ) \times
\begin{pmatrix} \bf \sigma_2 & 0 \\ 0 & \bf \sigma_2 \end{pmatrix}_{ij}\,. \nonumber
\eeq
So in the naive continuum limit for the diquark source term we have 
two copies of fundamental fermions.
Thus, one can expect that the partition function (\ref{z2}) after rooting procedure corresponds to QC$_2$D with $N_f = 2$ fundamental fermions.
Moreover, for sufficiently small lattice spacing $a$ the $\beta$-function of the theory (\ref{z2}) corresponds
to the $\beta$-function of QC$_2$D with two fundamental flavors (see below). For these reasons we believe, that
the partition function (\ref{z2}) in the continuum limit describes QC$_2$D with $N_f=2$ fundamental fermions.

\subsection{Observables}

In our simulations we measured the following observables:
\begin{itemize}
\item The Polyakov loop:
\begin{equation}
\label{eq:Pol_loop}
\< L \> = \frac{1}{N_s^3} \sum_{x_1,x_2,x_3 = 0}^{N_s - 1} \frac 1 2  \< \tr \prod_{x_4 = 0}^{N_\tau - 1} U_{x, 4} \>\,;
\end{equation}
\item The 
time-like Wilson loop around a rectangular contour  $C=R \times T$:
\begin{equation}
\label{eq:Wils_loop}
W(R,T) = \< \tr \biggr [ \prod_{C} U_{x, \mu} \biggl ] \> \,;
\end{equation}
\item The chiral condensate:
\begin{equation}
\label{eq:chiral_condensate}
a^3\<\bar q  q\> = a^3\<\bar q_{i \alpha}  q_{i \alpha}\> = - \frac{1}{N_s^3 N_\tau}\frac{\partial (ln\,Z)}{\partial (ma)}\,;
\end{equation}
\item The baryon density:
\begin{equation}
\label{eq:bar_number}
a^3 n_B = a^3 \frac{1}{2} \<\bar q_{i \alpha} \gamma_0 q_{i \alpha}\> = \frac 1 2 \frac{1}{N^3_s N_\tau} \frac{\partial (ln\,Z)}{\dmu}\,;
\end{equation}
\item The diquark condensate:
\begin{equation}
\label{eq:diquark_condensate}
a^3\<qq\>= - \frac{1}{N_s^3 N_\tau} \frac{\partial (ln\,Z)}{\partial \lambda} = a^3\<q^T_{i \alpha} \hat C \gamma_5 (\tau_2)_{ij} (\sigma_2)_{\alpha\beta} q_{j \beta}\> \,.
\end{equation}
\end{itemize}
In formulae (\ref{eq:chiral_condensate})--(\ref{eq:diquark_condensate}) the 
fields $\bar q, q$ are quark fields in the continuum theory,
$\hat C$ is the matrix of charge conjugation, $\tau_2$ and $\sigma_2$ are flavor 
and colour Pauli matrices, respectively. 
The quark fields have Dirac (not shown for the sake of brevity), colour ($\alpha$, $\beta$) and flavor indices 
($i,j$). 
Summation over repeated indices is tacitly understood. Notice that in addition to the quark contribution 
there is similar antiquark contribution to equation (\ref{eq:diquark_condensate}), which is not shown. 
This is because we work with positive chemical potential and in this region antiquark contribution to the equation (\ref{eq:diquark_condensate}) is exponentially suppressed.
In numerical calculations of the diquark condensate we have taken into account both quark and antiquark contributions.

The Polyakov and Wilson loops are meant to be sensitive to an eventual 
confinement/deconfinement phase transition. The chiral 
condensate is sensitive to breaking/restoration of the chiral symmetry. 
The diquark condensate is an order parameter for the transition to a 
phase, where scalar diquarks are condensed.

\subsection{Details of the simulation}

To study the phase diagram of QC$_2$D with $N_f=2$ flavors we used a 
$16^3 \times 32$ lattice, simulating with $\beta = 2.15$ and $ma = 0.005$,
what corresponds to a fixed temperature $T \approx 55$ MeV, lattice spacing
$a=0.112(1)~\mbox{fm}$, pion mass $M_{\pi}=378(4)~\mbox{MeV}$ and $m_{\pi} L_s \approx 3.4$ (see the
section~\ref{sec:scale_setting}). 
The simulation was carried out for a set of values of the chemical potential 
$\mu$ spanning the region $\mu \in [0 ; 1759]$ MeV ($\mua \in [0.0 ; 1.0]$). 
For each value of $\mu$ in the region $\mu \in [0 ; 1055]$ MeV ($\mua \in [0.0 ; 0.6]$)
we carried out the simulation at three values of the diquark source 
$\lambda=0.001$, $0.00075$ and $0.0005$. 
The measured data have been then extrapolated to $\lambda=0$. 
In the vicinity of the phase transition from the hadronic phase to the 
BEC phase $\mu = 176$, $211$, $246$ MeV ($\mua = 0.1$, $0.12$, $0.14$ respectively) we carried out simulations
at five values of the diquark source:
$\lambda = 0.001$, $0.000875$, $0.00075$, $0.000625$ and $0.0005$.
Simulations with higher $\mu$ are more computationally demanding,
thus for $\mu > 1055$ MeV ($\mua > 0.6$) only one value of the diquark source,
$\lambda=0.0005$, was used.

In the simulations we used the RHMC algorithm~\cite{Clark:2006wq, Clark:2006fx}. The 
fourth root in the action evaluation was approximated with the accuracy $\sim O(10^{-15})$.
For each pair of $\mu$ and $\lambda$ we generated 1000 --- 1500 MD trajectories 
after thermalization and performed measurements of the Polyakov loop (\ref{eq:Pol_loop})
at each trajectory and of the fermionic observables (\ref{eq:chiral_condensate})--(\ref{eq:diquark_condensate})
at each 10th trajectory. We employed the stochastic estimation
technique with Gaussian random sources to calculate fermionic traces
and used 100 --- 250 Gaussian random vectors per trace.

It is worth to mention, that we carried out a check of our simulation 
program through the comparison of our results with the  
QC$_2$D results existing in the literature.
In particular, we compared with the results of simulation 
of staggered fermions without rooting and chemical potential~\cite{Ilgenfritz:2012fw},
of staggered $N_f=2$ flavors and with $\mu=0$~\cite{Scheffler:2013jaa}, 
and of staggered $N_f=4$ flavors with non-zero chemical potential and non-zero diquark source~\cite{Kogut:2002cm}. 
For all these cases we found good agreement. 

\subsection{Scale setting and pion mass}
\label{sec:scale_setting}

First we performed additional measurements at zero values of the baryon 
chemical potential $\mu$ in order to calculate the $\beta$-function and the pion mass,
because the behaviour of the $\beta$-function provides a natural check for the correct continuum limit.
In these simulations we used a lattice with the size $16^3 \times 32$ as well.
To fix the physical scale, we extracted the heavy quark potential from smeared Wilson loops (1 HYP smearing~\cite{Hasenfratz:2001hp} step for temporal links was employed followed by 20 APE smearing~\cite{Albanese:1987ds} steps for spatial links, the
details are described in \cite{Bornyakov:2005iy}). From this potential
we extracted the Sommer scale parameter $r_0$. Assuming, that it is equal
to the Sommer scale parameter in real QCD, $r_0 = 0.468(4)$ fm in physical
units~\cite{Bazavov:2011nk}, we determined the lattice spacing. 

\begin{figure}[t]
\begin{center}
\includegraphics[scale=0.65, angle=0]{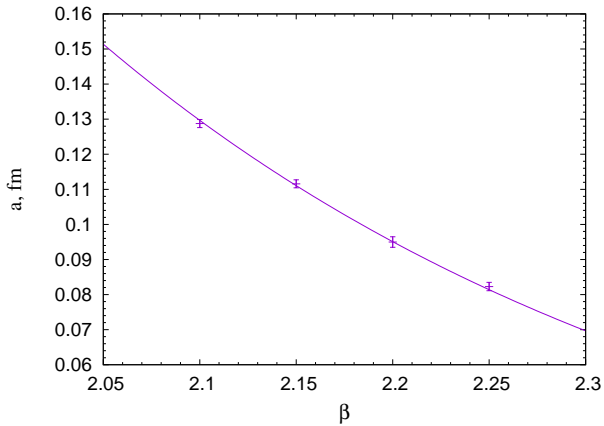}
\caption{(Color online) The dependence of the lattice spacing on the inverse coupling 
$\beta=4/g^2$.}
\label{figscale:}
\end{center}
\end{figure}

To carry out the scale setting we fixed the quark mass $ma=0.005$, the diquark source
$\lambda = 0.0005$ and varied $\beta \in [2.1 ; 2.25]$. 4000 MD trajectories were generated
for each value of $\beta$, measurements were performed at every 10th trajectory. The results
of the simulation are presented in Tab.~\ref{tabular:pion_masses} and in Fig.~\ref{figscale:}.

\begin{table}[tbp]
\centering
\begin{tabular}{|c|c|c|c|}
\hline
$\beta$ & $a, fm$ & $M_{\pi}, MeV$ \\ 
\hline
2.1  & 0.129(1) & 329(3)  \\
2.15 & 0.112(1) & 377(4)  \\
2.2  & 0.095(2) & 493(8) \\
2.25 & 0.082(1) & 561(9) \\
\hline
\end{tabular}
\caption{The lattice spacing $a$ and the pion mass $m_{\pi}$ for various 
values of the inverse coupling $\beta$ for the bare quark mass $ma = 0.005$ and $\lambda = 0.0005$.}
\label{tabular:pion_masses}
\end{table}

We found that for all considered values of $\beta$ the dependence of the lattice spacing $a$ can be reasonably fitted with the two-loop formula with $N_c = 2$ and $N_f = 2$:
\begin{equation}
\label{eq:beta_function}
\begin{split}
a(\beta) &= \frac{1}{\Lambda_L} \Bigl( \frac{4 \beta_0}{\beta} \Bigr)^{- \frac{\beta_1}{2 \beta_0^2}} exp \Bigl( \frac{- \beta}{8 \beta_0} \Bigr)\,,
\\
\beta_0 &= \frac{3}{8 \pi^2}\,, \qquad \beta_1 = \frac{29}{256 \pi^4}\,,
\end{split}
\end{equation}
with $\chi^2/{dof} = 0.47$ and $\Lambda_L = 0.0222(1)\,\text{fm}^{-1}$~\footnote{Scale setting for
$N_f = 4$ theory was performed in~\cite{Ilgenfritz:2012fw} on the lattice of the same size and with $ma = 0.01$,
obtained $\Lambda_L$ is nearly three times smaller than ours.}.
Good agreement between our data and the formula~(\ref{eq:beta_function}) for the $\beta$-function provides an argument that the partition function (\ref{z2}) in the continuum limit describes QC$_2$D with $N_f=2$ fundamental fermions.

To measure the pion masses we calculated the pion propagators $C_{\pi}(t,\vec{q}=0)$
for the same parameter sets, which were used for the scale setting. From the fit of
the pion propagators with the usual $\cosh$-form: $C_{\pi}(t,\vec{q}=0)=C\cosh(-m_{\pi}(t-T/2))$,
we extracted the pion masses, which are also presented in Tab.~\ref{tabular:pion_masses}
in physical units. We also checked that the results for Wilson loops and pion masses are practically independent of the value of the diquark source. For instance, at $\beta = 2.15$ for the $\lambda=0.0$ the pion mass is $m_{\pi}=378(4)$, for the $\lambda=0.0005$ the pion mass is $m_{\pi}=377(4)$ MeV and for the $\lambda=0.001$ the pion mass is $m_{\pi}=382(4)$ MeV.
We would like to note, that the pion mass in our study is smaller compared to previous studies~\cite{Kogut:2001if, Kogut:2001na, Kogut:2002cm, Hands:2006ve, Hands:2010gd, Cotter:2012mb, Boz:2013rca}.
\section{Numerical results}
\subsection{The diquark condensate}

In this section we are going to study the diquark condensate. 
It was noted above, that in the region $\mu \in [0.0 ; 1055]$ MeV 
($\mua \in [0.0 ; 0.6]$) the condensate is calculated for three values of the 
diquark source: $\lambda=0.0005$, $0.00075$, and $0.001$. To extrapolate our results to 
$\lambda \to 0$ we used a linear fit
\footnote{Notice that one can use the other fitting functions to extrapolate 
our results to $\lambda=0$. However, this will not change main results of this paper.}
of the data for all values of the 
chemical potential under investigation. The linear fit
turned out to be good
($\chi^2/dof \sim 1$) in the region $\mu \leq 141$ MeV ($\mua \leq 0.08$)
and $\mu \geq 263$ MeV ($\mua \geq 0.15$). 
For the values $\mua = 176$ MeV, $211$ MeV
and $246$ MeV ($\mua = 0.1$, $0.12$, $0.14$, respectively) 
a linear fit does not describe the data well.
We believe, that this fact
can be explained by the closeness of  
these $\mu$ values to the critical chemical potential $\mu^c$, 
where the system undergoes 
the phase transition from the hadronic phase to the phase with $\langle qq \rangle \neq 0$.

\begin{figure}[t]
\begin{center}
\includegraphics[scale=0.325, angle=0]{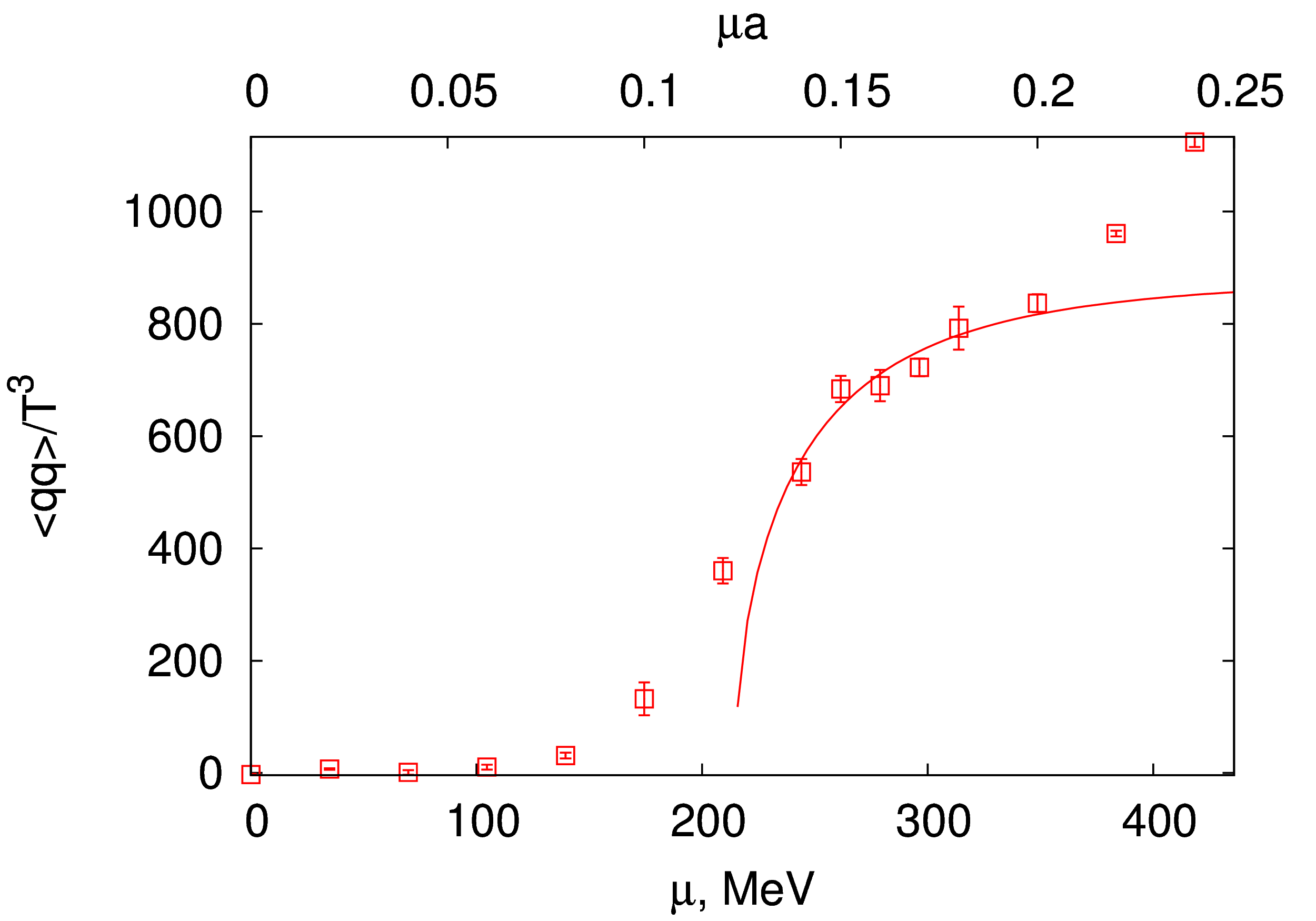}
\caption{(Color online) The diquark condensate $\langle qq \rangle/T^3$ as a function 
of $\mu$. 
The chemical potential is expressed in physical units (lower scale) and in 
lattice units (upper scale). The data are fitted by function (\ref{ChPTqq}).}
\label{fig2}
\end{center}
\end{figure}

In Fig.~\ref{fig2} we plot the diquark condensate $\langle qq \rangle$ 
(obtained by linear extrapolation to $\lambda=0$)
as a function of $\mu$ in the region $\mu \in [0.0 ; 440]$ MeV 
($\mua \in [0.0 ; 0.25]$).
It may be seen, that for $\mu \leq 141$ MeV ($\mua \leq 0.08$) the diquark condensate
$\langle qq \rangle$ is compatible with zero. However, for $\mu \geq 176$ MeV ($\mua \geq 0.1$)
the diquark condensate starts to deviate from zero. If we are sufficiently far 
from the position of the phase transition, one can try to use ChPT to describe the 
data~\cite{Kogut:1999iv, Splittorff:2000mm, Kogut:2000ek, Splittorff:2001fy}. 
In particular, ChPT predicts that the transition from the hadronic phase to the 
phase with $\langle qq \rangle \neq 0$ takes place at $\mu^c = m_{\pi} / 2$, and the 
behaviour of the diquark condensate above the transition would be given by 
the formula:
\beq
\langle qq \rangle = \langle \bar qq \rangle_0 \sqrt {1 - \biggr( \frac {\mu^c } {\mu} \biggl )^4},
\label{ChPTqq}
\eeq
where $\langle \bar qq \rangle_0$ is the chiral condensate at zero chemical 
potential. If one uses formula (\ref{ChPTqq}) to fit our data in the region $\mu \in [263 ; 352]$ MeV 
($\mua \in [0.15 ; 0.20]$), one gets $\mu^c = 215(10)$ MeV ($a\mu^c = 0.122(6)$) with $\chi^2/dof = 2.5$.
We plot the function (\ref{ChPTqq}) in Fig.~\ref{fig2}. 

One can try to fit the data by another function. To build it we recall that
in the ChPT the diquark condensate can be determined from the equation 
$\langle qq \rangle = \sqrt  {\langle \bar qq \rangle_0^2 - \langle \bar qq \rangle^2}$. 
In the ChPT for the $\mu>\mu^c$ the chiral condensate $\langle \bar qq \rangle$ drops with the chemical potential as $\sim 1/\mu^2$ and thus one gets (\ref{ChPTqq}). However, our data show (see below), that 
the chiral condensate drops slower: $\langle \bar qq \rangle \sim 1/\mu^{\alpha}$ with $\alpha=0.78(2)$. 
Thus it is reasonable to fit the data by the formula 
\beq
\langle qq \rangle = \langle \bar qq \rangle_0 \sqrt {1 - \biggr( \frac {\mu^c } {\mu} \biggl )^{2 \alpha}},
\label{ChPTqq1}
\eeq
with the power $\alpha$ mentioned above. The fit of the data by formula (\ref{ChPTqq1})
in the region $\mu \in [263 ; 352]$ MeV ($\mu \in [0.15, 0.20]$) gives $\mu^c = 193(10)$ MeV
($a\mu^c=0.110(6)$) with $\chi^2/dof = 1.4$. 

From these examples one sees, that the position of the critical point determined from the fitting 
procedure strongly depends on the fitting function.  
Nevertheless, one can state, that the results for $\mu^c$ 
are in reasonable agreement with ChPT.

\begin{figure}[t]
\begin{center}
\includegraphics[scale=0.35, angle=0]{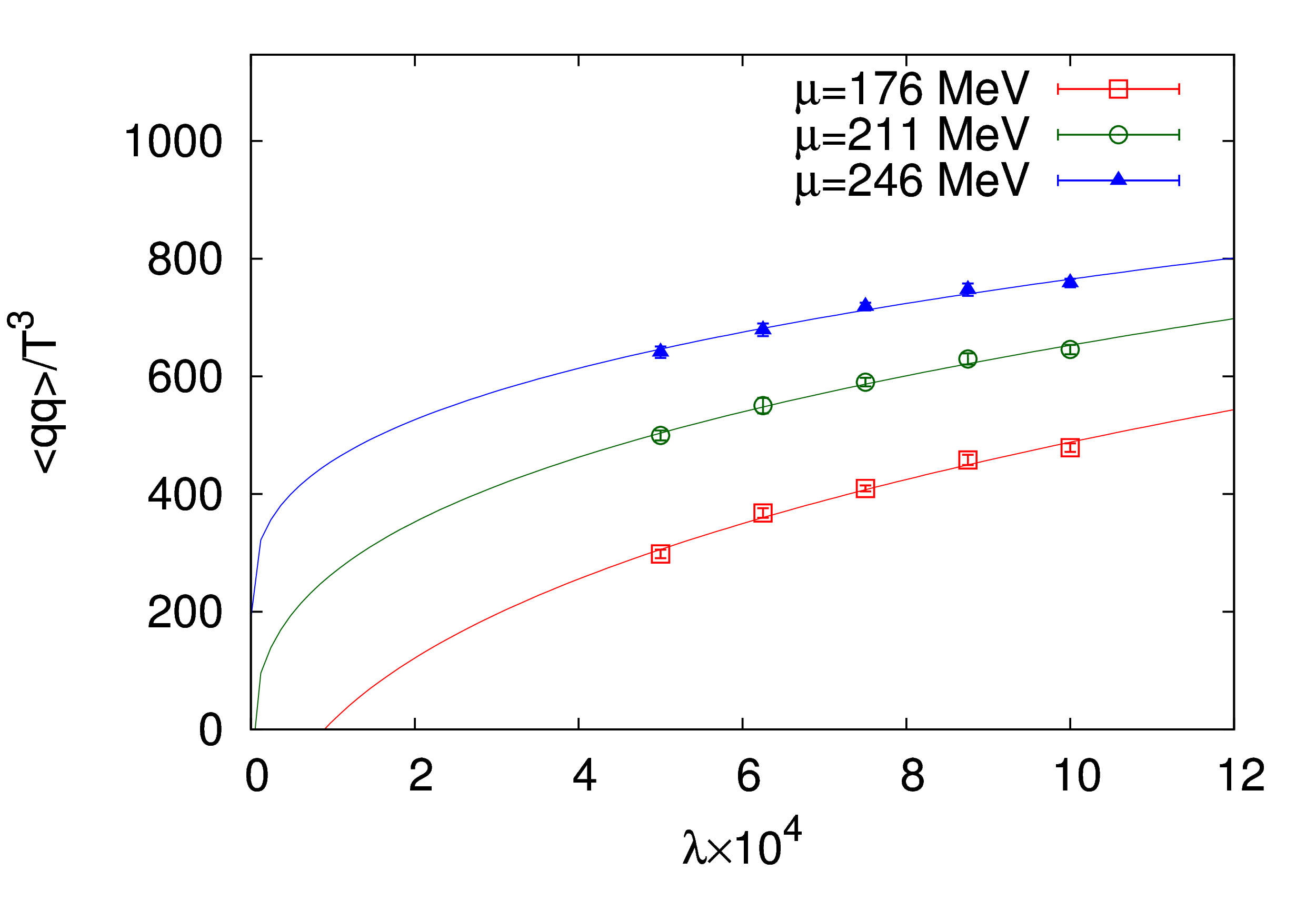}
\caption{(Color online) The diquark condensate $\langle qq \rangle/T^3$ as a function of 
$\lambda$ in the vicinity of the phase transition.}
\label{fig3}
\end{center}
\end{figure}

It is interesting to study the limit $\lambda \to 0$ of our data in the vicinity of 
the phase transition at $\mu = 176$, $210$ and $246$ MeV.
For these values of the chemical potential the diquark condensate was measured 
at five points $\lambda=0.0005$, $0.000625$, $0.00075$, $0.000825$ and $0.001$.
From ChPT we know, that at the critical chemical potential $\mu = \mu^c$ the behaviour
of the diquark condensate should be like $\langle qq \rangle \sim \lambda^{1/3}$.
Thus it is reasonable to fit the data in the vicinity of the phase transition
by the function $\langle qq \rangle = A + B \lambda^{1/3}$. The results of the fit are shown in Fig.~\ref{fig3}.
For all three values of the chemical potential the fit is good: $\chi^2/dof \sim 1$. We found, that for the smallest chemical potential value
$\mu = 176$ MeV ($\mua = 0.1$) the extrapolated value of the diquark condensate is negative:
$\langle qq \rangle|_{\lambda \to 0}=-0.012(2)$. Negative value of the condensate 
can be attributed to the fact that the value $\mu=176$ MeV is rather far 
from the critical point. For the next value $\mu = 211$
MeV ($\mua = 0.12$) the condensate is compatible to zero: $\langle qq \rangle|_{\lambda \to 0}=-0.0021(12)$.
Finally, for the largest value $\mu = 246$ MeV  ($\mua = 0.14$) the condensate is greater than zero:
$\langle qq \rangle|_{\lambda \to 0} = 0.0058(14)$. These results 
indicate that $\mu = 211$ MeV ($\mua \simeq 0.12$) is closer to the critical 
point than $\mu = 176$ MeV ($\mua = 0.1$) and $\mu = 246$ MeV ($\mua = 0.14$), 
which agrees within the uncertainty with the value of the critical point 
obtained above. 

To summarize: in the region $\mu < \mu^c$ the system is in the hadronic phase 
with zero diquark condensate. In the region $\mu > \mu_c$ the system is in the BEC phase
with nonzero diquark condensate. In the region $\mu \in [0.0 ; 352]$ MeV ($\mua \in [0.0 ; 0.20]$)
our results for the diquark condensate are in good agreement with ChPT predictions. From Fig.~\ref{fig2}
one sees, that in the region $\mu > 352$ MeV ($\mua > 0.2$) the data start to deviate from ChPT description.

\begin{figure}[t]
\begin{center}
\includegraphics[scale=0.325, angle=0]{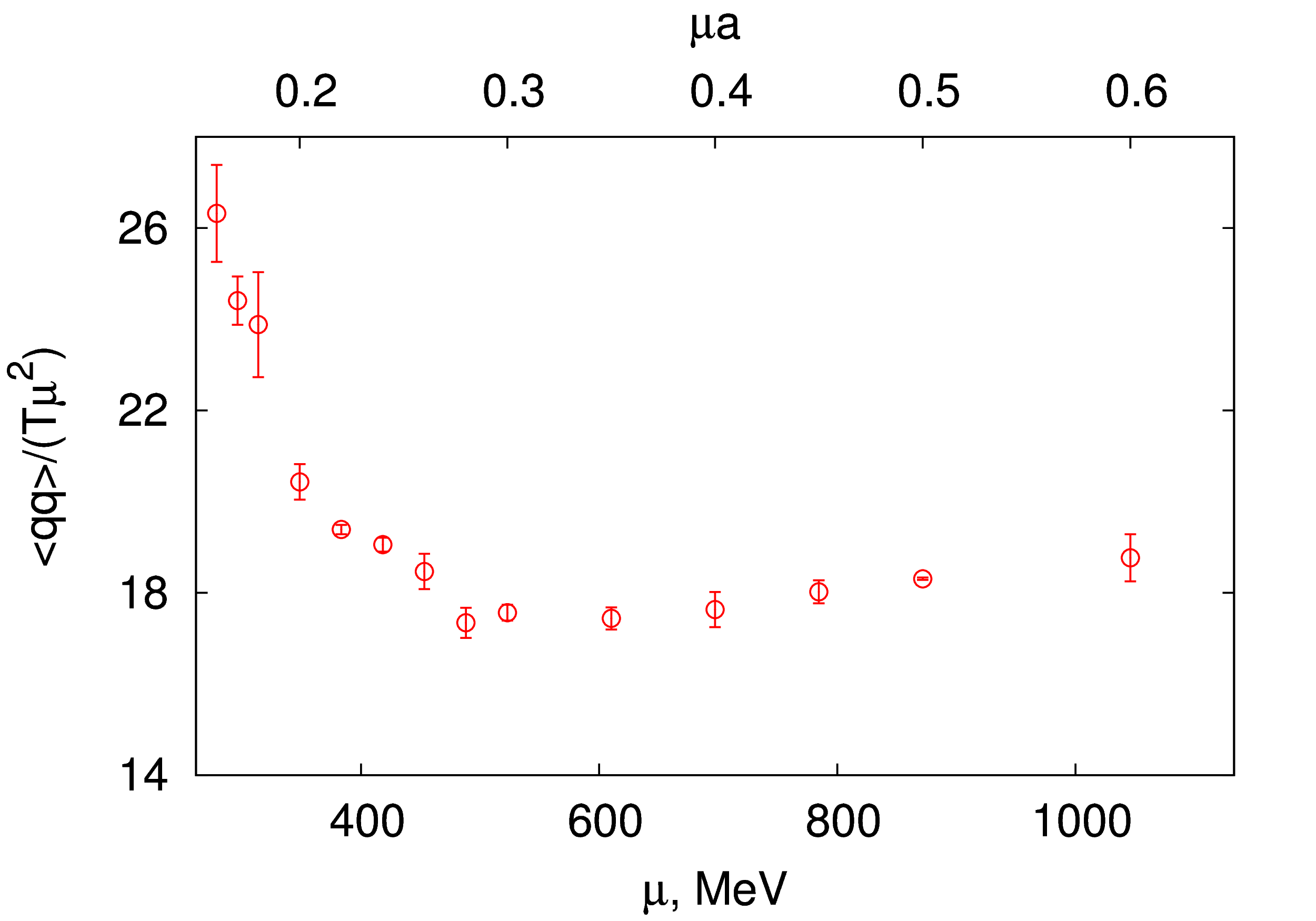}
\caption{(Color online) The ratio $\langle qq \rangle/(T\mu^2)$ as a function of $\mu$. 
The chemical potential is expressed in physical units (lower scale) and 
in lattice units (upper scale).}
\label{fig4}
\end{center}
\end{figure}

Let us consider the region of larger chemical potential 
$\mu > 352$ MeV ($\mua > 0.2$). To understand what happens
in this region, 
we plot in Fig.~\ref{fig4} the linearly extrapolated diquark 
condensate, divided by $T\mu^2$, as a function of $\mu$.
As visible from this plot, in 
the region $\mu \in [528 ; 1055]$ MeV ($\mua \in [0.3 ; 0.6]$) 
there is a plateau, {\it i.e.} the
value of the diquark condensate is proportional to the surface of a sphere with the radius $\mu$: 
$\langle qq \rangle \sim \mu^2$. This is a characteristic property of the BCS theory, where the
condensate appears on the Fermi surface and where it is proportional to the density of states
on this surface. Thus we conclude, that for $\mu > 528$ MeV ($\mua > 0.3$) the system reveals
properties of the BCS phase, and that the transition from the BEC to the BCS phase is smooth.

It is worth to note, that in~\cite{Kogut:2002cm} $N_f = 4$ theory was simulated on the $16^4$ lattice
at $\beta = 1.85$ with $ma = 0.05$ and the BCS phase has not been observed. 
According to~\cite{Ilgenfritz:2012fw}, the lattice spacing for this set of parameters is larger than lattice spacing in our simulations. In our study of the critical chemical potential is $\mu^c a \simeq 0.12$, 
whereas in~\cite{Kogut:2002cm} it was found that $\mu^c a \simeq 0.29$. From the relation $\mu^c = m_\pi/2$ one might conclude that in~\cite{Kogut:2002cm} the pion is more than two times heavier than in our simulations. This remarkable physical difference may be the reason why in the previous studies with $N_f=4$ the BCS phase has not been realized.

\begin{figure}[t]
\begin{center}
\includegraphics[scale=0.325, angle=0]{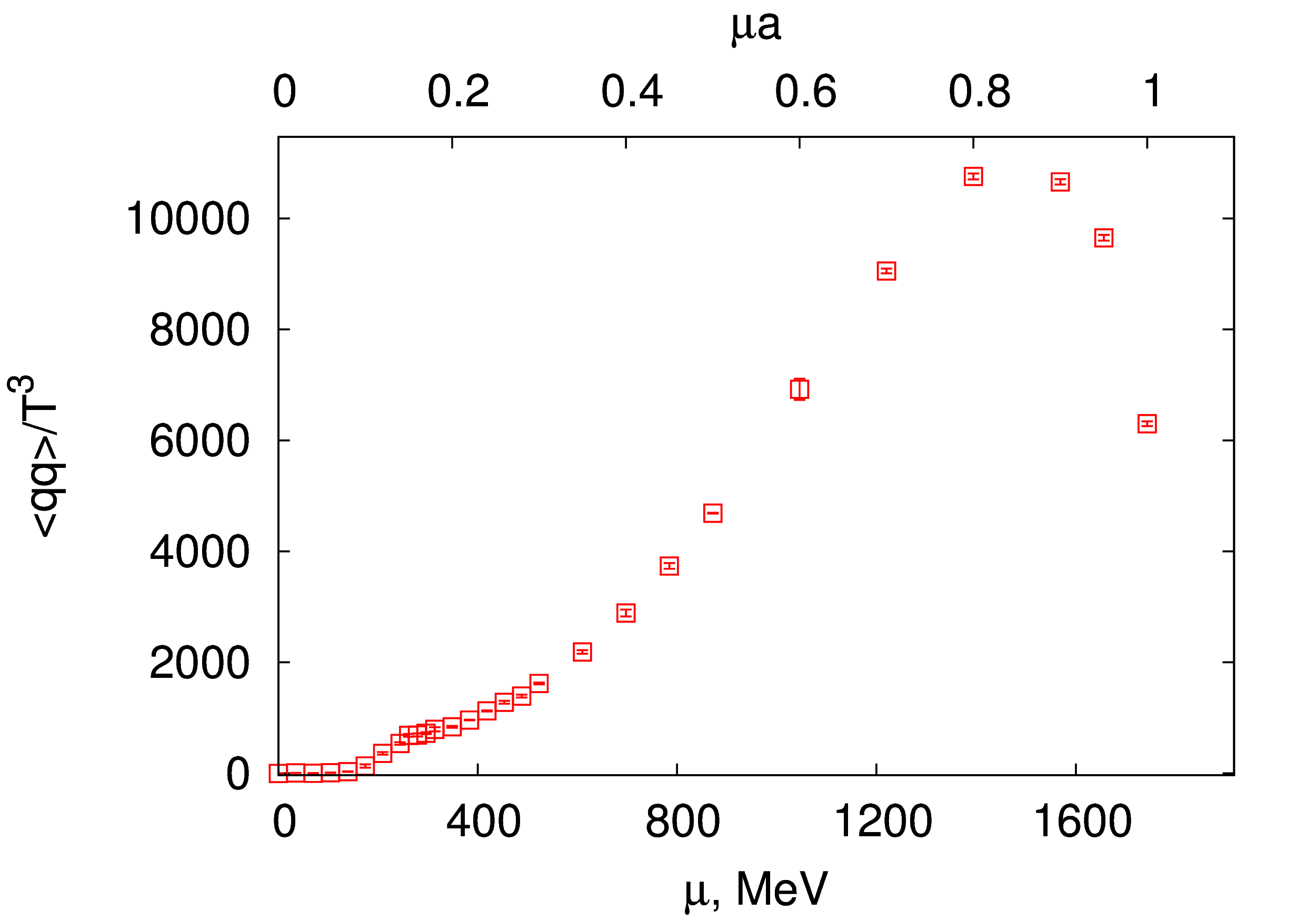}
\caption{(Color online) The diquark condensate $\langle qq \rangle/T^3$ as a function of 
$\mu$. The chemical potential is expressed in physical units (lower scale) and in lattice units (upper scale).}
\label{fig5}
\end{center}
\end{figure}

In the region $\mu > 1055$ MeV ($\mua > 0.6$) the simulations become very computationally demanding.
At the same time in this region the value of the diquark condensate becomes less sensitive to the value
of the source $\lambda$, compared to the BEC phase. We believe that this might be related to 
the fact that the larger the $\mu$ the larger the fermion mass gap, 
which plays a role of the regulator of the fermion determinant.
For this reason for $\mu > 1055$ MeV ($\mua > 0.6$) we used $\langle qq \rangle|_{\lambda=0.0005}$ as the estimate of the value of the condensate at $\lambda = 0$. In Fig.~\ref{fig5} we plot the diquark condensate  $\langle qq \rangle$ as a function of $\mu$ throughout the whole region under study. In the region $\mu > 1055$ MeV ($a\mu > 0.6$) the condensate starts to deviate from the BCS behaviour, and after $\mu > 1410$ MeV ($a\mu > 0.8$) the condensate decreases. Such a descent of the diquark condensate $\langle qq \rangle$ in the region $\mua \sim 1$ has already been observed in refs.~\cite{Kogut:2001na,Kogut:2002cm}. This behaviour might be connected with a saturation effect, and therefore can be considered as a lattice artifact.

\begin{figure}[t]
\begin{center}
\includegraphics[scale=0.325, angle=0]{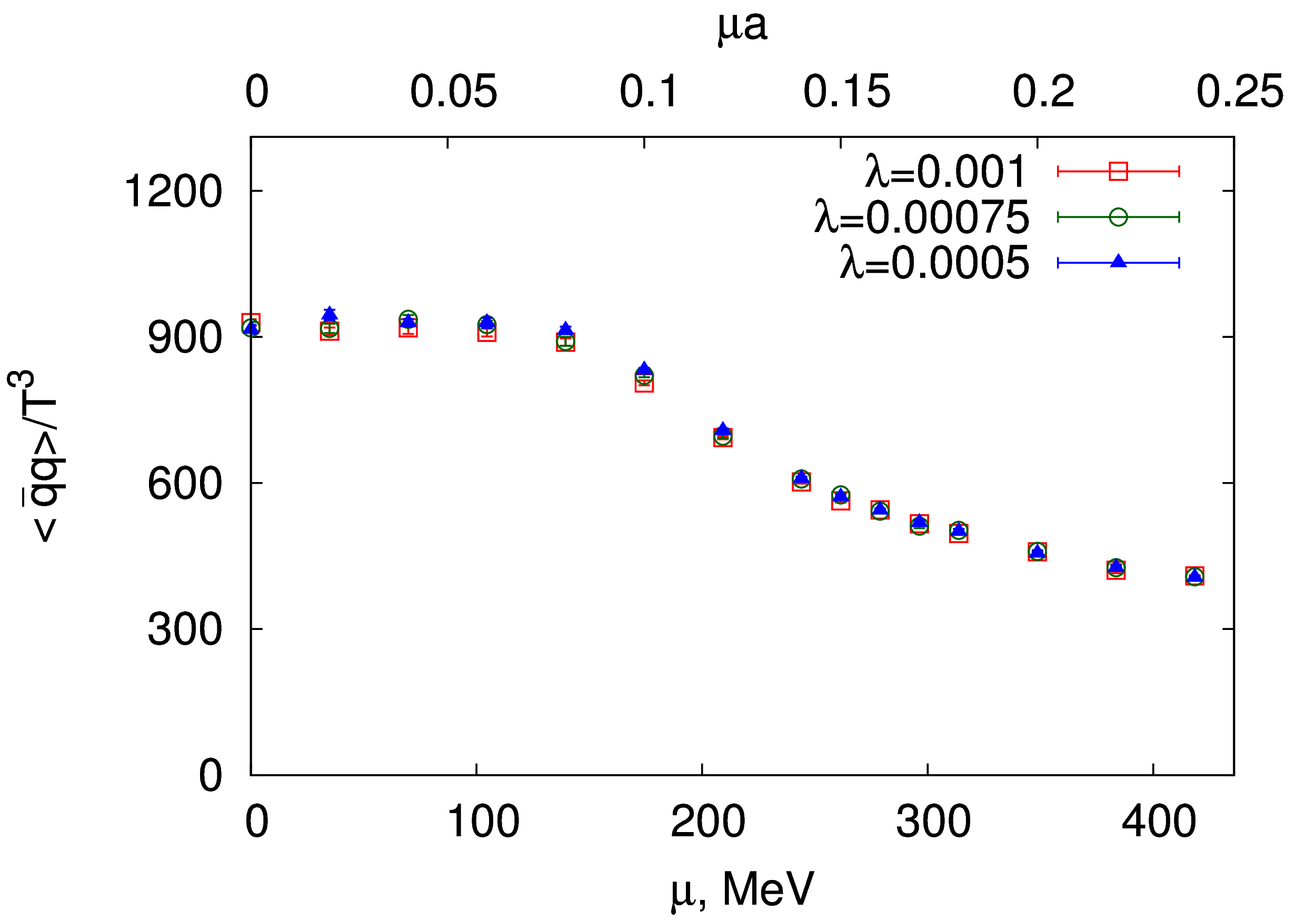}
\caption{(Color online) The chiral condensate $\langle \bar qq \rangle/T^3$ as a function 
of $\mua$ for the values $\lambda=0.001$, $0.00075$ and $0.0005$ of the diquark source. 
The chemical potential is expressed in physical units (lower scale) and 
in lattice units (upper scale).}
\label{fig6}
\end{center}
\end{figure}

\subsection{The chiral condensate}

\begin{figure}[t]
\begin{center}
\includegraphics[scale=0.325, angle=0]{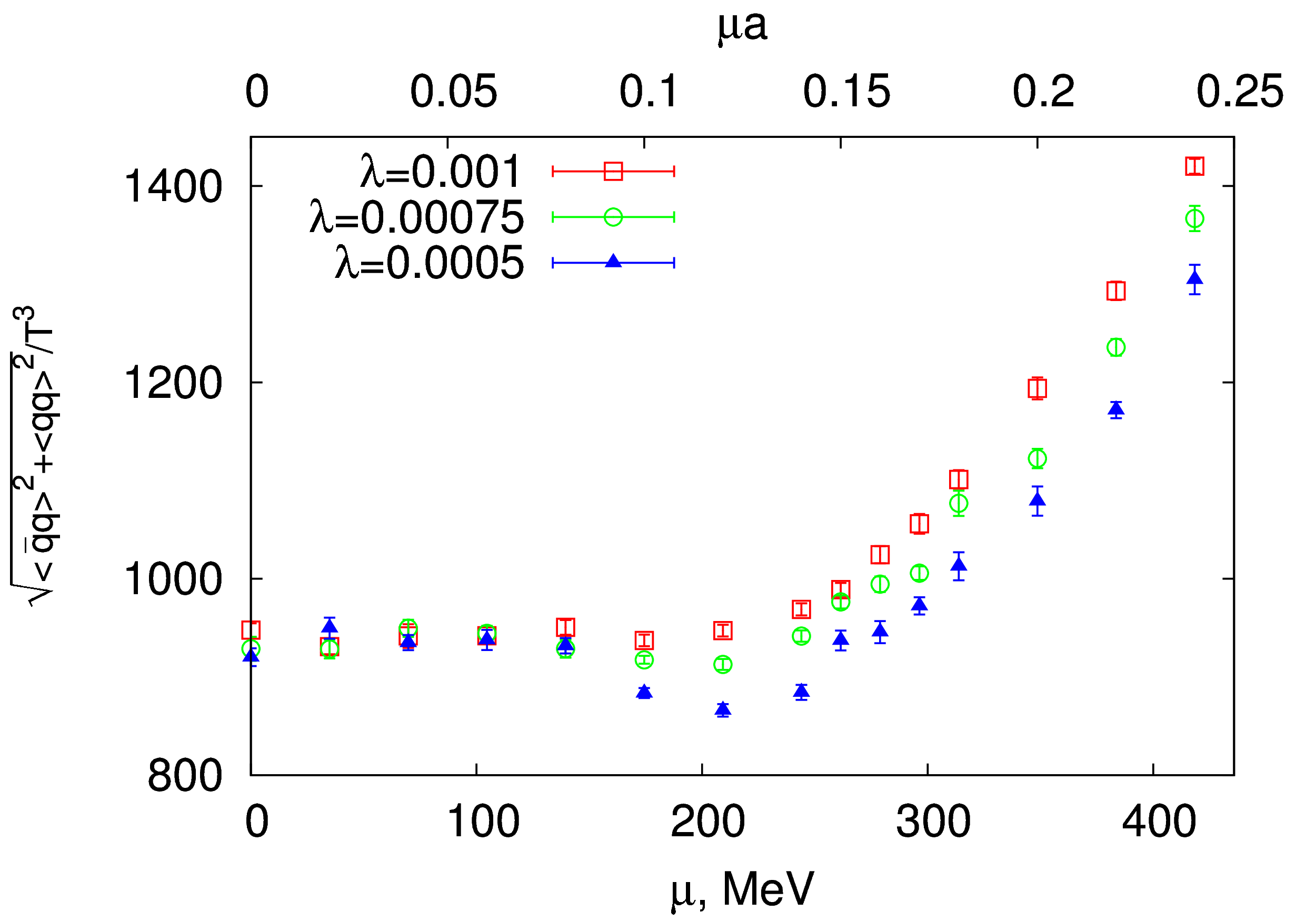}
\caption{(Color online) The combination 
$\sqrt {\langle qq \rangle^2 + \langle \bar qq \rangle^2}/T^3$ 
of diquark and chiral condensates as a function of $\mu$. The chemical potential
is expressed in physical units (lower scale) and in lattice units (upper scale).}
\label{fig7}
\end{center}
\end{figure}

Next let us consider the chiral condensate $\langle \bar qq \rangle$.
In Fig.~\ref{fig6} we plot the chiral condensate as a function of $\mu$ 
in the region $\mu \in [0.0 ; 440]$ MeV ($\mua \in [0.0 ; 0.25]$) for the
following three values of the diquark 
source: $\lambda = 0.001$, $0.00075$ and $0.0005$. From the Fig.~\ref{fig6} 
it is obvious, that the dependence of the chiral condensate 
on the source $\lambda$ is very weak. Except for a few fluctuations, the values
of the chiral condensate $\langle \bar qq \rangle$, calculated at different values of
$\lambda$, are equal to each other within the uncertainties. The next observation is
that up to $\mu < 176$ MeV ($\mua < 0.1$) the chiral condensate does not depend on the
chemical potential. In the region $\mu > 176$ MeV, where the system is in the vicinity of the 
transition to the BEC phase, the chiral condensate starts to decrease. These properties
are in agreement with ChPT predictions 
(see Figures 4 and 5 in paper \cite{Kogut:2000ek}).
An interesting prediction of ChPT is that 
in the whole region, where ChPT is applicable, 
a relation between the chiral condensate and the diquark condensate holds:
$\langle qq \rangle^2 + \langle \bar qq \rangle^2 = const$~\cite{Kogut:1999iv}.
Note that this ``circle relation'' is valid only in the leading order approximation,
and it is violated by the next-to-leading order corrections~\cite{Splittorff:2001fy}.
Our lattice results allow us to address the question how well this relation 
is satisfied. In Fig.~\ref{fig7} we plot the combination
$\sqrt {\langle qq \rangle^2 + \langle \bar qq \rangle^2}$ as a function of 
$\mu$.
From this plot one sees, that for the diquark source $\lambda=0.001$ this 
relation is well satisfied up to $\mu < 263$ MeV ($\mua < 0.15$). 
For bigger $\mu$ values one can see the deviation from the ``circle relation'' 
for all values of the diquark source $\lambda$ under consideration.
Note also that for the values $\lambda=0.00075$ and $0.0005$, which move the 
system closer to the phase transition, the deviation from the ``circle 
relation'' is clearly seen in the transition region $\mu \in [176 ; 246]$ MeV 
($\mua \in [0.1 ; 0.14]$). The smaller the source $\lambda$ is chosen,
the larger is the deviation. We believe, that the deviation of our results
from the relation 
$\sqrt {\langle qq \rangle^2 + \langle \bar qq \rangle^2}=const$
in the region $\mu \in [176 ; 246]$ MeV  can be explained by the closeness of the system
to the transition point, 
where a mean field study of ChPT is not applicable.

\begin{figure}[t]
\begin{center}
\includegraphics[scale=0.325, angle=0]{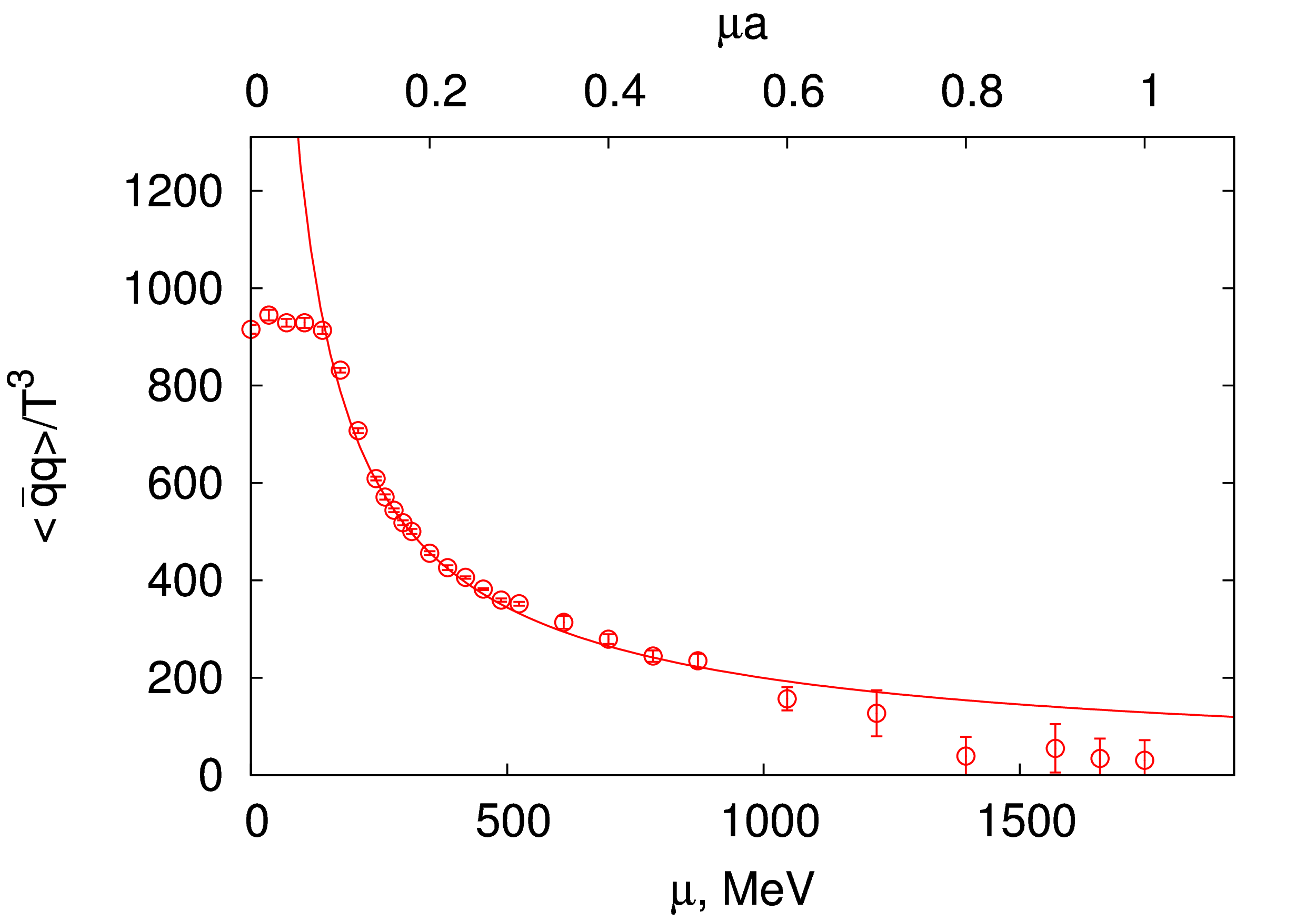}
\caption{(Color online) The chiral condensate $\langle \bar qq \rangle/T^3$ as a function
of $\mu$. The chemical potential is expressed in physical units (lower scale) and in lattice units (upper scale).}
\label{fig8}
\end{center}
\end{figure}

Now let us consider the chiral condensate throughout the full region $\mu \in [0 ; 1759]$ MeV
($\mua \in [0.0 ; 1.0]$). In Fig.~\ref{fig8} we plot the chiral condensate calculated for the 
smallest diquark source value $\lambda = 0.0005$ as a function of $\mu$. It was noted above that
the chiral condensate is practically insensitive to the values of $\lambda$, thus the value of
the chiral condensate at $\lambda = 0.0005$ can be taken as the value at $\lambda = 0$.
According to ChPT, at $\mu > \mu^c$ the chiral condensate drops as
\beq
\langle \bar q q  \rangle = \langle \bar q q  \rangle_0 \biggr ( \frac {\mu^c} {\mu} \biggl )^2,
\eeq
where $\langle \bar q q \rangle_0$ denotes the chiral condensate at zero 
chemical potential. To check this prediction in the region $\mu \in [263 ; 352]$ MeV
($\mua \in [0.15 ; 0.20]$) we fit our data by a power law
$\langle \bar q q  \rangle = A/\mu^{\alpha}$. This ansatz fits our data
well ($\chi^2/dof = 0.3$) with the exponent $\alpha=0.78(2)$. It is interesting to note, that this fit gives 
a satisfactory description of the data up to  $\mu \sim 1055$ MeV ( $\mua \sim 0.6$). Thus, one sees that the chiral
condensate drops slower with increasing chemical potential than ChPT predicts. Similar slower decrease
of the form $ \langle \bar{q} q \rangle \sim 1 / \mu$ was observed in~\cite{Cotter:2012mb} on the gauge
ensembles generated with $N_f = 2$ Wilson quarks.

Good agreement with the LO ChPT prediction for the chiral condensate dependence on the chemical potential was found in~\cite{Hands:2000ei} for $N_f=1$ adjoint flavor, and in~\cite{Kogut:2001na} for $N_f=4$ fundamental flavors, the latter study being carried out at $\beta=1.5$. On the other hand, in~\cite{Kogut:2002cm} another lattice study at $\beta= 1.85$ has been performed, where the chiral condensate was found to decrease as $\langle \bar{q} q \rangle \sim 1 / \mu^\alpha$ with $\alpha = 1 \cdots 1.3$ depending on the $\lambda$ value (see Table 3 of the Ref.~\cite{Kogut:2002cm}). The same dependence of the $\langle \bar q q \rangle$ on the baryon chemical potential was also observed in~\cite{Hands:1999md} for $N_f = 8$ fundamental flavors at $\beta = 1.3$. We conjecture that the behaviour of the chiral condensate is rather sensitive on the coupling regime of the theory. If $\beta$ is small enough and the system is in the strong coupling regime the leading order of ChPT is sufficient, and higher order effects are weak.

\begin{figure}[t]
\begin{center}
\includegraphics[scale=0.325, angle=0]{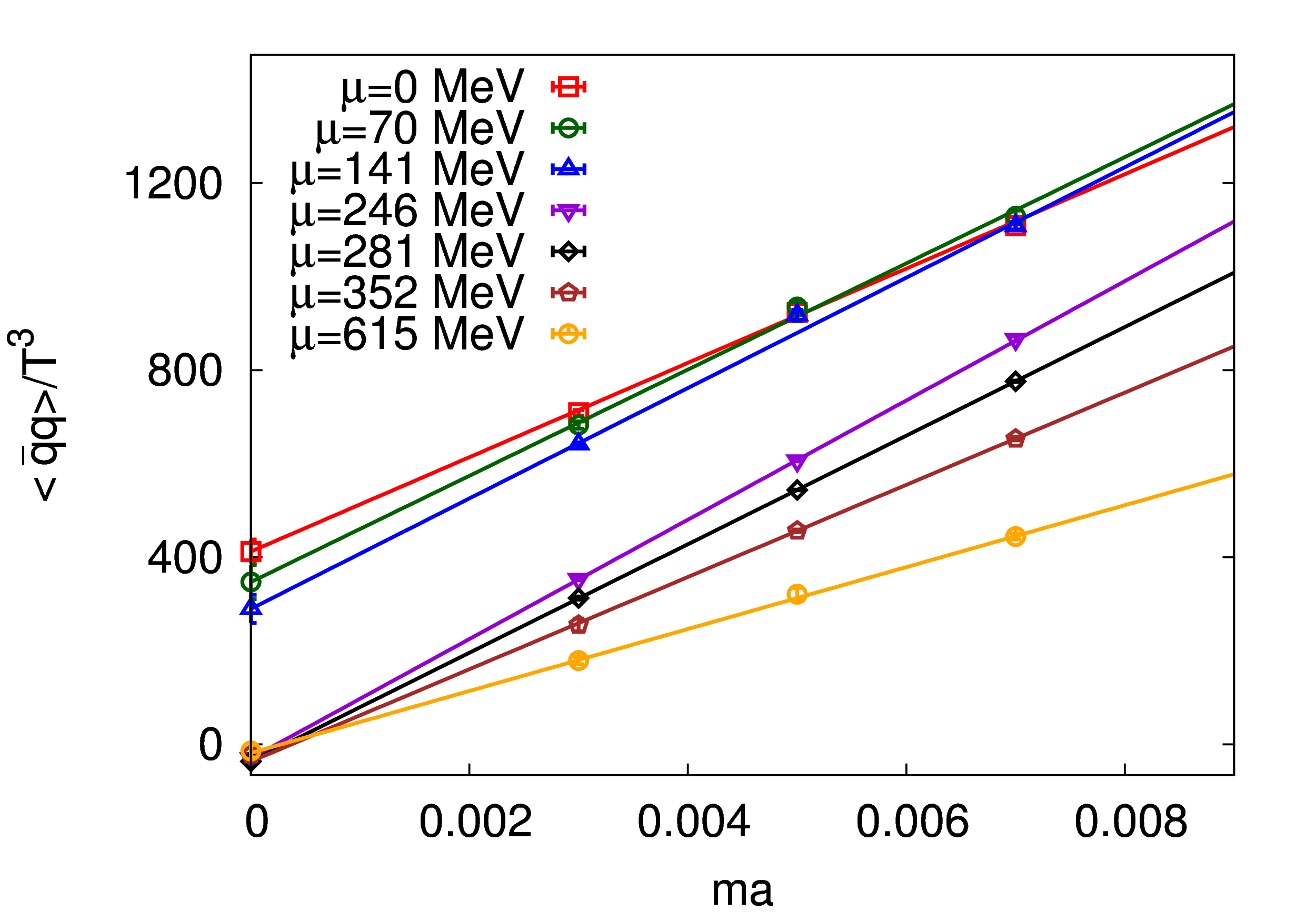}
\caption{(Color online) The chiral limit of the chiral condensate 
$\langle \bar qq \rangle/T^3$, taken for different values of the chemical 
potential. The quark mass is expressed in lattice units.}
\label{fig9}
\end{center}
\end{figure}

Finally, it is interesting to study the chiral symmetry breaking in the chiral 
limit for different regions of the chemical potential. In Fig.~\ref{fig9} we plot
the chiral condensate for different values of the chemical potential as function of the quark mass. 
As an example we took a few values of the chemical potential in the hadronic 
phase: $\mu=0$, $70$ and $141$ MeV ($\mua = 0.0$, $0.04$, $0.08$, respectively), in the BEC phase: $\mu=246, 281, 352$ MeV 
($\mua = 0.14, 0.16, 0.20$, respectively), and in the BCS phase: $\mu = 615$ MeV ($\mua = 0.35$).
At these fixed values of the chemical potential we linearly extrapolate our data to $ma = 0$. 
It is seen from Fig.~\ref{fig9}, that chiral symmetry breaking exists in  
the chiral limit within the hadronic phase (values $\mu=0$, $70$ and $141$ MeV),
whereas there is no chiral symmetry breaking in the chiral limit in the BCS phase ($\mu = 615$ MeV). 
We also found, that the chiral limit of the chiral condensate at the points 
$\mu=246, 281, 352$ MeV (in the BEC phase) are vanishing, but it is difficult 
to claim, that there is no chiral symmetry breaking in the whole BEC phase:
when we take the chiral limit, we change the pion mass and thus shift the 
critical point closer $\mu^c$ to zero. This effect is not important for the values of the 
chemical potential sufficiently far from the phase transition, but it might be important
close to the phase transition. Note, that the absence of chiral symmetry breaking in the
chiral limit within the BEC phase agrees with ChPT predictions. 

\subsection{The baryon density}

In this section we are going to consider the baryon number density $n_B$. It clear 
from formulae (\ref{z2}) and (\ref{eq:bar_number}), that the baryon density 
depends on the square of the diquark source, $\lambda^2$, but not on $\lambda$.
Thus, to get the baryonic density at zero diquark source, it is reasonable to 
fit our data for each value of the chemical potential
by an ansatz $n_B(\lambda) = A + B \lambda^2$.

\begin{figure}[t]
\begin{center}
\includegraphics[scale=0.325, angle=0]{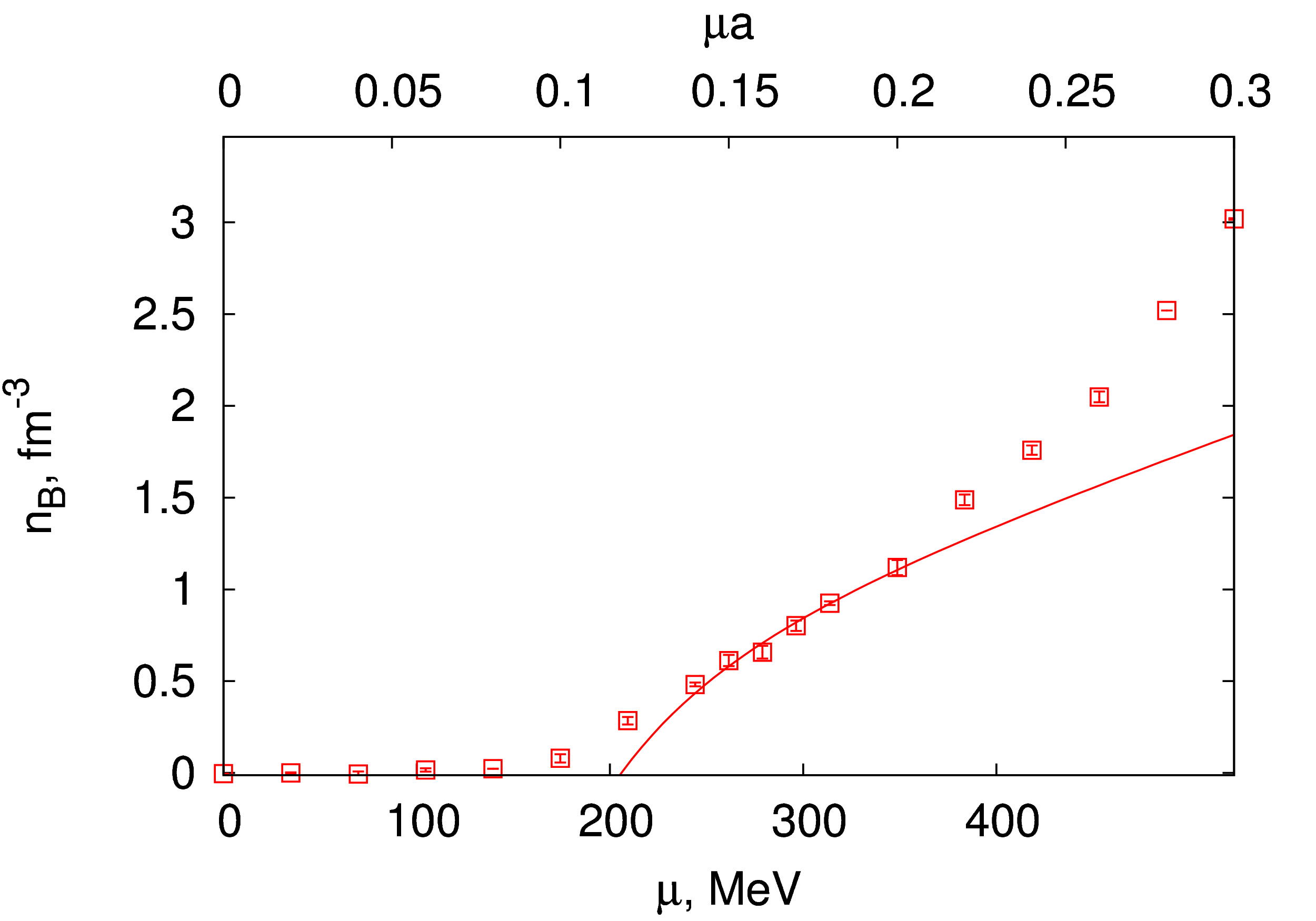}
\caption{(Color online) The baryon density $n_B$ in physical units,
as a function of $\mu$. The chemical potential
is expressed in physical units (lower scale) and in 
lattice units (upper scale).}
\label{fig10}
\end{center}
\end{figure}

In Fig.~\ref{fig10} we plot the baryon density in the region 
$\mu \in [0.0 ; 528]$ MeV  ($\mua \in [0.0 ; 0.3]$).
It is clear, that for all $\mu < 176$ MeV ($\mua < 0.1$) the baryon density
is vanishing within the uncertainty of the calculation. In the vicinity
of the phase transition ($\mu \geq 176$ MeV) the baryon density starts
to deviate from zero, and for larger values of the chemical potential 
it rises with increasing $\mu$. ChPT predicts, that the dependence of
the baryon density on the chemical potential above $\mu^c$ is given by a formula
$n_B \sim \mu - \mu_c^4/\mu^3$. In the region $\mu \in [263 ; 352]$ MeV 
($\mua \in [0.15 ; 0.20]$) we fit our data by this formula in order to extract
the critical chemical potential $\mu^c$. The fit is of good quality, $\chi^2/dof = 1.2$,
and the result is $\mu^c = 207(7)$ MeV ($a\mu^c = 0.118(4)$). This value is in agreement
with our previous results for $\mu^c$, obtained from the $\langle qq \rangle$ fits. 
From Fig.~\ref{fig10} it is visible, that for bigger chemical potential, $\mu > 352$ MeV
($\mua > 0.2$), our data deviate from the ChPT prediction.

\begin{figure}[t]
\begin{center}
\includegraphics[scale=0.325, angle=0]{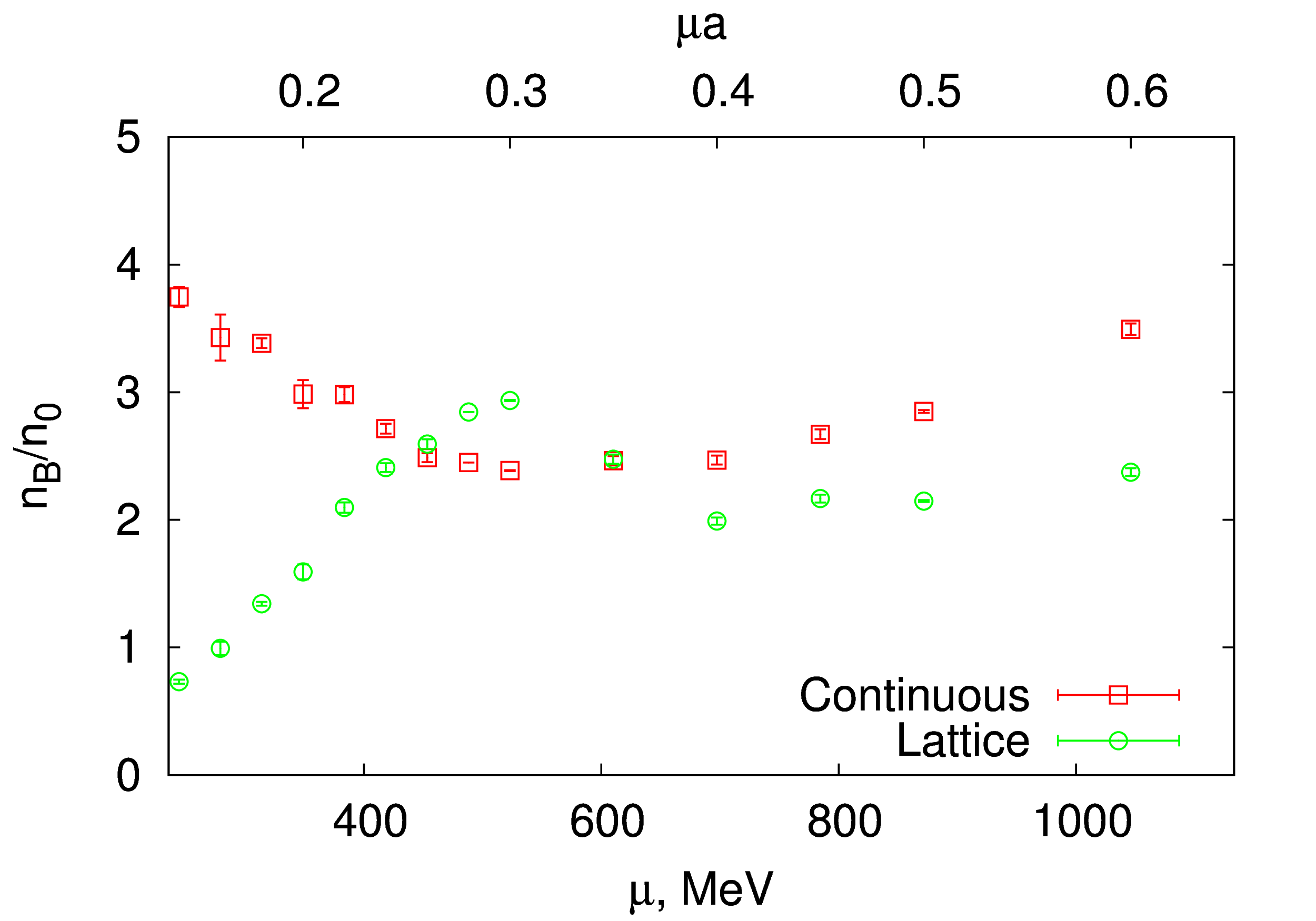}
\caption{(Color online) The ratio $n_B/n_0$ as a function of the chemical potential $\mu$.
For the square points, the reference density $n_0$ denotes the baryon density for
{\it free continuum fermions}, $n_0 = (2\mu^3) / (3\pi^2)$, whereas for the circle points
the reference density $n_0$ denotes the baryon density for {\it free lattice fermions}.
The chemical potential is expressed in physical units (lower scale) and in lattice units (upper scale).}
\label{fig11}
\end{center}
\end{figure}

Next, let us consider the baryon density at even larger values of the chemical 
potential. In Fig.~\ref{fig11} we plot the ratio $n_B/n_0$ as a function of 
$\mu$, where for the square points the reference density $n_0$ is the baryon
density for free continuum fermions at $T = 0$, $n_0 = (2 \mu^3) / (3 \pi^2)$,
and for the circle points $n_0$ is the baryon number density for free lattice fermions.
It can be seen, that in the region $\mu \in [528 ; 1055]$ MeV ($\mua \in [0.3 ; 0.6]$) these ratios
are slowly varying functions of the chemical potential, taking values in the region $2.0 \ldots 2.5$,
whereas the measured baryon density changes by an order of magnitude. We believe, that the scaling
of the baryon density $n_B \sim n_0$ confirms the conclusion that in the region $\mu \in [528 ; 1055]$ MeV
the system is in a BCS-like phase. The relevant degrees of freedom in this phase are quarks, which mostly
live inside the Fermi sphere with a condensate of Cooper pairs on the Fermi surface. At the same time,
the fact that $n_B/n_0 \sim 2.0 \ldots 2.5$, but not $\sim 1.0$, can be attributed to UV and IR effects.
Similar effects on the baryon density, although of smaller size ($n_B / n_0 \sim 1.1 \ldots 1.5$), 
were observed in~\cite{Cotter:2012mb} (see also the Fig. 6 therein for the demonstration of UV and
IR artifacts in $n_B$).

\subsection{The gluon observables}

In this section we study the gluon observables 
Polyakov loop (\ref{eq:Pol_loop})
and
Wilson loops (\ref{eq:Wils_loop}). Similarly to the chiral condensate the gluon
observables are not sensitive to the value of the $\lambda$, thus we take these observables
calculated at the smallest value $\lambda = 0.0005$ as their values at the $\lambda = 0$.

We measured the average of the Polyakov loop as a function of the chemical potential.
The result of this measurement is that for all values of the chemical potential studied in 
this paper the average Polyakov loop is vanishing within the uncertainty of the calculation.

\begin{figure}[t]
\begin{center}
\includegraphics[scale=0.325, angle=0]{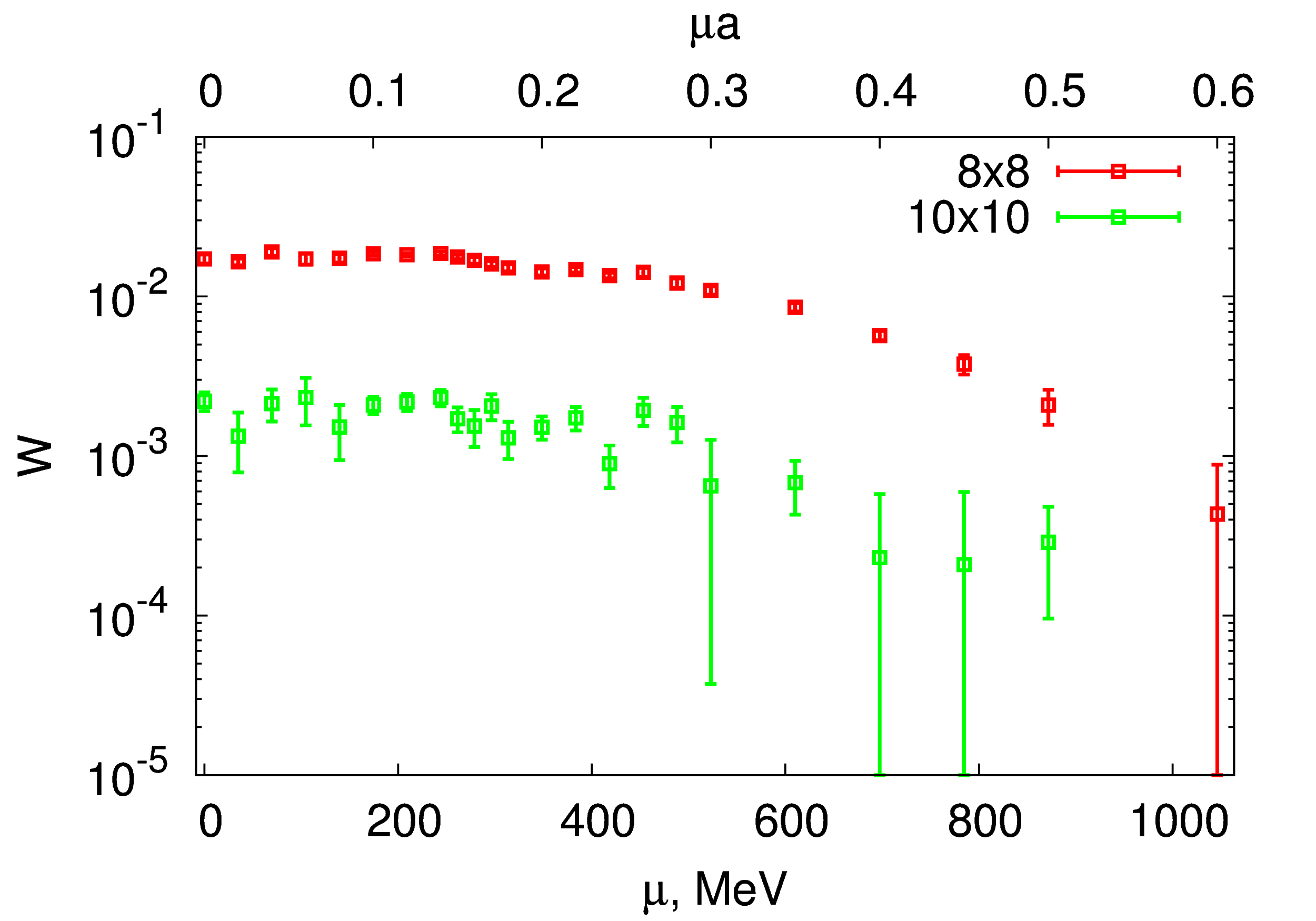}
\caption{(Color online) The time-like Wilson loops for the contours $8 \times 8$ and $10 \times 10$ as a functions of the chemical potential $\mu$.}
\label{fig12}
\end{center}
\end{figure}

Furthermore, in order to investigate the confinement properties of the system, we have calculated time-like
Wilson loops (\ref{eq:Wils_loop}) for the 
quadratic contours of the size $8 \times 8$ and $10 \times 10$ (for  larger
Wilson loops we obtained 
results compatible with zero) as functions of the chemical potential. 
The same smearing
strategy, as discussed in the section II.D, 
was employed for these Wilson loops measurements. The results are shown
in Fig. \ref{fig12}. 
One learns from this plot, that for $\mu > 352$ MeV ($\mu > 0.2$) the 
Wilson loops
decrease with the growth of the chemical potential. At small $\mu$,
for  $\mu \in [0 ; 263]$ MeV ($\mua \in [0.0 ; 0.15]$),
a plateau for both Wilson loops may be noticed. 
From these results one can conclude, that the system is in a confined phase 
for all values of the chemical potential under consideration. The possible
explanation for this behaviour may be the absence of the Debye screening in
two-color QCD at zero temperature~\cite{Kanazawa:2009ks, Rischke:2000qz}.

\section{Discussion and conclusion}

In conclusion, in this paper 
we have carried out a low-temperature scan of the phase diagram of dense 
two-color QCD
with $N_f = 2$ quarks. 
The study has been conducted
 using lattice simulations with rooted staggered quarks.

Our results can be summarized as follows. At small chemical potential
$\mu < \mu^c=m_{\pi}/2 \sim 200$ MeV we observe a {\bf hadronic phase}. 
In this 
phase QC$_2$D matter is in confinement, chiral symmetry is broken, 
the diquark 
condensate (\ref{eq:diquark_condensate}) vanishes and the baryon number
density is also zero. Relevant degrees of freedom in this phase are Goldstone bosons.

In the region $\mu^c < \mu < \mu^d \sim 352$ MeV we observe {\bf the BEC phase}.
Characteristic feature of this phase is Bose-Einstein 
condensation of scalar diquarks. The order parameter
for the transition to the BEC phase is the diquark condensate, which develops a non-zero value in the region
$\mu > \mu^c$. Within the uncertainty of the calculation $\mu_c = m_{\pi} / 2$, where $m_{\pi}$ is the pion mass
at zero chemical potential. In this phase, QC$_2$D matter has also confining properties, whereas
the baryon density is non-zero.
Based on our detailed results for the 
onset of the diquark condensate we believe, that the transition from the 
hadronic
to the BEC phase should be of the second order. Relevant degrees of freedom in the BEC phase are Goldstone bosons as well. 

We have also found, that the chiral limits of the chiral condensate at the points 
$\mu = 246,\,281,\,352$ MeV  in the BEC phase are vanishing.
Nevertheless, it is difficult to claim, that there is no chiral symmetry breaking
in the whole BEC phase, since when we take the chiral limit -- we change the pion mass
and thus shift the critical point $\mu^c$. This effect is not important for 
the values of the 
chemical potential sufficiently far from the phase transition, but it might be 
essential close to the phase transition.

It is important to notice, that for all values of the chemical potential 
$\mu < \mu^d$ our results are in good agreement with the predictions 
of ChPT. 
An exception is the chiral condensate, which drops with increasing chemical 
potential slower than ChPT predicts in leading order. 
This behaviour of the chiral condensate
might be explained by  higher radiative corrections.

In the region $\mu > \mu^d$ our data start to deviate from ChPT predictions. 
The physical origin of this deviation can be understood as follows. 
At $\mu = \mu^d$ the baryon number density is $n_B \sim 1~\mbox{fm}^{-3}$ 
(see Fig.~\ref{fig10}). 
In SU(3) theory, a baryon 
density $n_B \sim 1~\mbox{fm}^{-3}$
is of the order, when a gas of baryons can not be considered anymore as dilute. 
The interactions of baryons 
at a density $n_B > 1~\mbox{fm}^{-3}$ play an important role 
and can not be taken into account as a perturbation, as it is done within ChPT. 
On the contrary, for a density $n_B < 1~\mbox{fm}^{-3}$ a gas of baryons
can be considered as dilute and ChPT is applicable. From this consideration
one may conclude, that QC$_2$D in the region $\mu^c < \mu < \mu^d$ is an analog
of the dilute baryon gas of SU(3) QCD. It is remarkable, that the density
at which one expects the transition from dilute gas to dense baryon matter in the SU(3) QCD
is very close to that in the QC$_2$D.

If we further increase the chemical potential, starting from 
$\mu \sim 500 - 600$ MeV, one can observe that the diquark condensate scales 
as $\langle qq \rangle \propto \mu^2$ and the baryon density scales as 
$n_B \propto \mu^3$. Physically, this implies that the relevant degrees of 
freedom are quarks, which are mostly living inside the Fermi sphere with a condensate 
of Cooper pairs on the Fermi surface. These properties are clear hints in favor 
of the {\bf BCS phase}. In this phase the chiral symmetry is restored in the 
chiral limit.
Our measurements of the time-like Wilson loops imply, that the system  
still keeps the confinement property in this phase. 
In addition our data confirm, that the transition from the BEC to the BCS
phase is smooth, 
if there is a phase transition at all.

The BCS phase extends up to $\mu \sim 1000 - 1100$ MeV. In the region 
$\mu \in [1100 ; 1410]$ MeV the ratio $\langle qq \rangle / \mu^2$
drops, the baryon density scaling is $n_B \sim \mu^3$, the chiral condensate 
is very small, and the system is 
still retaining the confinement property.
It is not quite clear, what happens
in this region, but most likely we are facing with lattice artifacts, related with the fact,
that $\mua$ is close to $1$. 
In the region $\mu > 1410$ MeV ($\mua > 0.8$) the diquark condensate
begins to drop, and $n_B$ is close to saturation. 

The results obtained in this paper are in reasonable agreement with the 
results of Refs.~\cite{Kogut:2001if, Kogut:2001na, Kogut:2002cm}. In these 
papers the authors studied the phase diagram of QC$_2$D with $N_f = 4$ flavors
of staggered fermions. What concerns a low temperature scan of the phase 
diagram, these authors observed the succession of a hadronic phase and
the BEC phase, with their properties well described by ChPT, but they didn't
find a BCS phase.

In Refs.~\cite{Hands:2006ve, Hands:2010gd, Cotter:2012mb, Boz:2013rca} 
the QC$_2$D phase diagram with $N_f = 2$ flavors was studied through lattice
simulation with Wilson fermions. In a low temperature scan of the phase diagram
the authors observed a hadronic phase, followed by the BCS phase with deconfinement. 
Probably, the BEC phase has been missed in their simulations 
due to the violation of chiral symmetry by Wilson fermions. 
In addition these authors observed the transition to the deconfinement phase 
at $\mu \sim 800$ MeV for a
temperature $T = 47$ MeV~\cite{Cotter:2012mb}. 
In our study we don't observe the transition
to the deconfinement phase up to the chemical potential $\mu \sim 1410$ MeV.
In order to understand the origin of the disagreement between our results and the results
of the other groups one should carry out more numerical simulations with different set
of lattice parameters, but with the same $N_f$ and at the same physical point.

It is interesting to mention the results of Ref.~\cite{McLerran:2007qj}, where the phase diagram
of SU($N_c$) QCD was studied in the limit $N_c \to \infty$. The authors of this paper predicted
the following phases: firstly, a hadronic phase is observed at sufficiently small chemical potential;
when the chemical potential reaches $\mu = m_N / N_c$ the baryonic density 
ceases to vanish,
and there starts a phase of a dilute nuclear gas, which is similar to the BEC phase of the
QC$_2$D theory. Further enhancing the chemical potential, 
this study has ended with the so-called ``quarkyonic phase''. 
In this phase there is a Fermi sphere of quarks, at the surface of which 
baryons are living.
The system is in confinement, but chiral symmetry is restored. The described ``quarkyonic phase'' at
large $N_c$ may be similar to the BCS phase of the QC$_2$D theory. Using this physical picture it is not difficult to estimate the value of the chemical potential, where the quarkyonic phase becomes manifest. To do this we note, that the thickness of the surface
layer, where strong interactions are important, is $\sim \Lambda_{QCD}$. Then the ``quarkyonic phase''
becomes manifest, when the volume inside the Fermi sphere $\sim 4/3~\pi \mu^3$ becomes larger
than the volume of the surface layer, modified by strong interactions which is $\sim 4 \mu^2 \Lambda_{QCD}$. 
Thus we get $\mu > 3 \Lambda_{QCD}$. If we take $\Lambda \sim 200$ MeV, 
the ``quarkyonic phase'' starts at $\mu> 600$ MeV, 
what is in good agreement with 
the result of our present paper. One can also expect, that the value of the
chemical potential, where 
the ``quarkyonic phase'' starts in 
SU(3) theory is close to that in QC$_2$D, $\mu \sim 600$ MeV, since the 
$\Lambda_{QCD}$ values
in both theories are close to each other.

Finally, we summarize that in this paper we have carried out a low temperature 
scan of the phase diagram for the QC$_2$D theory with two flavors of quarks. 
We have shown that the phase structure of this theory has a lot of similarities
with SU($N_c$) theory at large $N_c$. Since the predictions of the SU($N_c$) 
theory at large $N_c$ start to work already at $N_c=2$, one can use QC$_2$D 
to make quantitative estimates for SU(3) QCD with chemical potential, 
which is directly inaccessible due to the sign problem.

\begin{acknowledgments}
Numerical simulations were performed at the supercomputer of Institute for Theoretical and Experimental Physics
(ITEP), at the federal center for collective usage at National Research Center (NRC) ``Kurchatov Institute''
(http://computing.kiae.ru/) and at Moscow State University (MSU) supercomputer ``Lomonosov''. The work of A.Y.K.
was supported by Russian Foundation for Basic Research (RFBR) Grant No. 16-32-00048 and Dynasty Foundation.
The work of V.V.B., A.V.M. and A.A.N., which consisted of developing the program for generation of gluon
configurations and studying the baryon density, chiral condensate and collecting statistics, was supported
by the Russian Science Foundation (RSF) grant under Contract No. 15-12-20008.
\end{acknowledgments}

\bibliography{bibl}

\end{document}